\documentclass[5p,twocolumn]{elsarticle}
\usepackage{graphicx,amssymb}
\usepackage{lineno,hyperref}
\usepackage{color}
\usepackage{mathtools}
\usepackage{upgreek}
\usepackage[utf8]{inputenc}

\makeatletter
    \def\ps@pprintTitle{%
       \let\@oddhead\@empty
       \let\@evenhead\@empty
       \let\@oddfoot\@empty
       \let\@evenfoot\@oddfoot
    }
\makeatother

\biboptions{sort&compress}

\Newlabel{zz}{$\dagger$}

\begin{document}
\begin{frontmatter}

\title{Correlated pion-proton pair emission off hot and dense QCD matter}

\author[5]{J.~Adamczewski-Musch}
\author[11,10]{O.~Arnold}
\author[9]{C.~Behnke}
\author[17]{A.~Belounnas}
\author[8]{A.~Belyaev}
\author[11,10]{J.C.~Berger-Chen}
\author[2]{A.~Blanco}
\author[9]{C.~Blume}
\author[11]{M.~B\"{o}hmer}
\author[2]{P.~Bordalo}

% ...
\address[2]{LIP-Laborat\'{o}rio de Instrumenta\c{c}\~{a}o e F\'{\i}sica Experimental de Part\'{\i}culas , 3004-516~Coimbra, Portugal}
% ...
\address[5]{GSI Helmholtzzentrum f\"{u}r Schwerionenforschung GmbH, 64291~Darmstadt, Germany}
% ...
\address[8]{Joint Institute of Nuclear Research, 141980~Dubna, Russia}
\address[9]{Institut f\"{u}r Kernphysik, Goethe-Universit\"{a}t, 60438 ~Frankfurt, Germany}
\address[10]{Excellence Cluster 'Origin and Structure of the Universe' , 85748~Garching, Germany}
\address[11]{Physik Department E62, Technische Universit\"{a}t M\"{u}nchen, 85748~Garching, Germany}
% ...
\address[17]{Laboratoire de Physique des 2 infinis Irène Joliot-Curie, Université Paris-Saclay, CNRS-IN2P3. , F-91405~Orsay , France}
% ...
\address[c]{also at Frederick University, 1036~Nicosia, Cyprus}
% ...

\begin{abstract}
In this letter we report the first multi-differential measurement of correlated pion-proton pairs from 2 billion Au+Au collisions at $\sqrt{s_{NN}}=2.42$~GeV collected with HADES. 
In this energy regime the population of $\Delta(1232)$ resonances plays an important role in the way energy is distributed between intrinsic excitation energy and kinetic energy of the hadrons in the fireball.
The triple differential ${\rm d^3}N/{\rm d}M_{\uppi^{\pm}p}{\rm d}p_{\rm T}{\rm d}y$ distributions of correlated $\uppi^{\pm}\text{p}$ pairs have been determined by subtracting 
the $\uppi$p combinatorial background using an iterative method. 
The invariant-mass distributions in the $\Delta(1232)$ mass region show strong deviations from a Breit-Wigner function with  vacuum width and mass.
The yield of correlated pion-proton pairs exhibits a complex isospin, rapidity and transverse-momentum dependence.
In the invariant mass range $1.1<M_{\text{inv}}$(GeV$/c^2) < 1.4$, the yield is found to be similar for $\uppi^+$p and $\uppi^-$p pairs, and to follow a power law  $\langle\mathrm{A}_{\mathrm{part}}\rangle^{\alpha}$, where $\langle\mathrm{A}_{\mathrm{part}}\rangle$ is the mean number of participating nucleons. 
The exponent $\alpha$ depends strongly on the pair transverse momentum ($p_\text{T}$) while its $p_\text{T}$-integrated and charge-averaged value is $\alpha=1.5\pm0.08^{\rm st}\pm0.2^{\rm sy}$.

\end{abstract}


\end{frontmatter}

\section{Introduction}

Understanding the structure and bulk properties of hot and dense QCD matter created in heavy-ion collisions is one of the most complex challenges in modern physics. 
Yields of particles, produced in heavy-ion collisions in a broad beam-energy range from Bevalac/SIS18 to the LHC, are spanning over several orders of magnitude and can be satisfactorily described with Statistical Hadronization Models~\cite{Hagedorn1965,Cabibbo:1975,Hagedorn1985,Cleymans1993,Cleymans2006_freeze-out,Jacak2012_Hot_Matter,Becattini:2016,Andronic_QCD_phase_nature:2017}. 
The thermodynamic parameters, temperature (T) and baryon chemical potential ($\mu_B$), extracted from the respective thermal fits show a smooth energy dependence which can be parametrized as shown in \cite{Cleymans2006_freeze-out,Andronic_QCD_phase_nature:2017}

The abundance of correlated $\uppi$N pairs at freeze-out grows with baryon-chemical potential. 
At vanishing values of $\mu_B$, where pions and other light meson states dominate, baryon and antibaryon resonances are equally important for modelling the yields and spectra of particles \cite{Markert_resonances_in_QGP:2008,Lo:2017ldt,PokManLo2017_effect_on_pt_spectra,Andronic2018_proton_puzzle}. 
At high values of $\mu_B$, reached in fixed-targed collisions at beam energies of few GeV per nucleon, excited baryons play an important role, so that the term resonance matter is sometimes used ~\cite{Metag:threshold_production_probe_resonance_matter_1992}. 
Strong modifications of the baryon spectral functions are predicted as a function of temperature and density~\cite{vanHeesRapp_In_medium_delta:2004}, which motivates attempts towards their direct reconstruction.
It was demonstrated that baryon resonances are  key ingredients in modelling the medium modification of vector meson spectral functions~\cite{Rapp:1997,Post_Leupold_in_medium_spectral_functions:2003,Wolf_eta_and_dilepton_prod_in_HIC:1993} and the production of strange particles below their free nucleon-nucleon production threshold~\cite{Oliinychenko_forced_thermalization:2016,Steinheimer_high_mass_resonances:2015,Li_hyperon_hyperon_scatt:2012,Zetenyi:2018oyd}. 
The role of baryon resonances in the production of dileptons and strange particles in heavy-ion collisions around 1\textit{A} GeV has also been addressed in  recent works~\cite{HadesNature2019,HADES_xim_arkcl:2009,HADES_delta_dalitz_2017,HADES_Lambda_k0_auau:2018,HADES_phi_kmin_auau:2017,Hades:ArKClPhi_2013,Hade:ArKCl_epem2011}. 
Therefore, a direct reconstruction of short-lived ($c\tau\sim 1~\text{fm}/c$) baryon resonances has the potential to further constrain different model calculations.

In heavy-ion collisions in the few GeV energy regime, nucleons can be excited in first chance collisions to $\Delta$ and N$^*$ resonances,
which can further interact in the reaction zone via rescattering or absorption processes.  
In addition, since the life time of $\Delta$ and N$^*$ resonances is shorter than the duration of the hot and dense phase, their decay products will engage in further generations of baryon resonances until the system reaches the kinetic freeze-out. 
The yield and the line-shape of resonances reconstructed using correlated $\uppi$p pairs are therefore sensitive to the complex dynamics of interacting  pions, nucleons and baryon resonances formed in the collisions. 

The S-matrix formulation of statistical mechanics offers a useful framework for describing interacting particles in thermal equilibrium~\cite{Dashen:1969}. 
Such calculations demonstrate that modeling of the hadron interactions (repulsive force between hadrons or excluded volume effect) is channel dependent and is provided by the respective phase shift. For example, the phase shift for P33 takes into account the $\Delta(1232)$ in the $\uppi$N channel~\cite{Weinhold_thermo_delta:1997,Lo:2017ldt}.

Detailed predictions for the the properties of baryon resonances produced in Au+Au reaction above 1\textit{A} GeV were provided by transport models \cite{Vogel_high_pt_probes:2009,Bass:probin_delta_matter_AuAu_1gev_1994,Bass:high_pt_pions_probes_1994,Reichert2019_UrQMD_Delta_For_HADES,Reichert:2020uxs}, which anticipated a continuous reduction of the most probable invariant mass of the $\Delta(1232)$ in the course of the collision, due to energy dissipation in the regeneration processes. 
Such effects could be experimentally scrutinized by investigating the transverse momentum ($p_{\text{T}}$) dependence of the correlated $\uppi$p pair mass ($M_{\text{inv}}$), as pairs characterized by large $p_{\text{T}}$ are expected to decouple earlier from the fireball~\cite{Bass:high_pt_pions_probes_1994}. 
The numbers of correlated $\uppi$N pairs in the different isospin configurations and their kinematics reflect the competition between the $\Delta$ decay, scattering or absorption,  which depends on the nucleon density and the scattering and absorption cross sections.
The study of $\uppi^{-}$p and $\uppi^{+}$p in the same experiment can therefore provide important complementary information.

First investigations of correlated pion-proton pair measurements were motivated by the observation of an apparent downward mass shift  of the $\Delta$ resonance in inclusive charge-exchange reactions \cite{Ellegaard_first_delta_shifts_PhysRevLett1983} attributed to a coherent excitation of resonance-hole states~\cite{Delorme_Guichon_delta_hole_1989}. 
However, in such peripheral reactions, the $\uppi$p pairs mostly exhibit properties of a quasi-free process \cite{Hennino:1992bw}. 
Significant modifications of the $\Delta(1232)$ line shape with respect to the vacuum distribution have been previously obtained in elementary~\cite{DIOGENE_Trzaska:1991,STAR_Resonances_Abelev:2008} and heavy-ion collisions~\cite{E814_Barrette_Si_Al_Pb:1994,Muntz1997,FOPI_NiNi_Hong:1997,EOS_Delta_NiCu_Hjort:1997,FOPI_Eskef:2001,medium_Pelte:1999}. 
In these studies, a substantial fraction of pions in the final state has been identified as originating from $\Delta$ decays. 
However, the previously collected data samples, based on $10^5$-$10^7$ events, were not sufficient to carry out multi-differential studies. 
One important challenge of such measurements in heavy systems as Au+Au is the large abundance of uncorrelated pions and protons which dominate the pair spectrum and has to be subtracted in order to access the true correlated signal. 
For such a purpose, an iterative technique, which improves the identification of the combinatorial background compared to the most commonly used mixed event-technique has been developed and applied in this work~\cite{Kornakov:iterative_2018}.

For the first time, we have measured the production of correlated pion--proton pairs from Au+Au collisions at $\sqrt{s_{NN}}=2.42$~GeV collected with HADES~\cite{Agakishiev:2009am} at the SIS18 synchrotron at GSI, Darmstadt. 
In this letter, we present the  $M_{\text{inv}}-p_{\text{T}}-$ rapidity $(y)$ distributions of correlated $\uppi^-$p and $\uppi^+$p pairs, in particular their $p_{\text{T}}$-dependent line-shape parameters, multiplicity per event as a function of event centrality as well as inverse slope parameters as a function of rapidity.

\section{Experiment and analysis method}

\begin{figure}[tb!]
\centering
\includegraphics[width=0.8\linewidth]{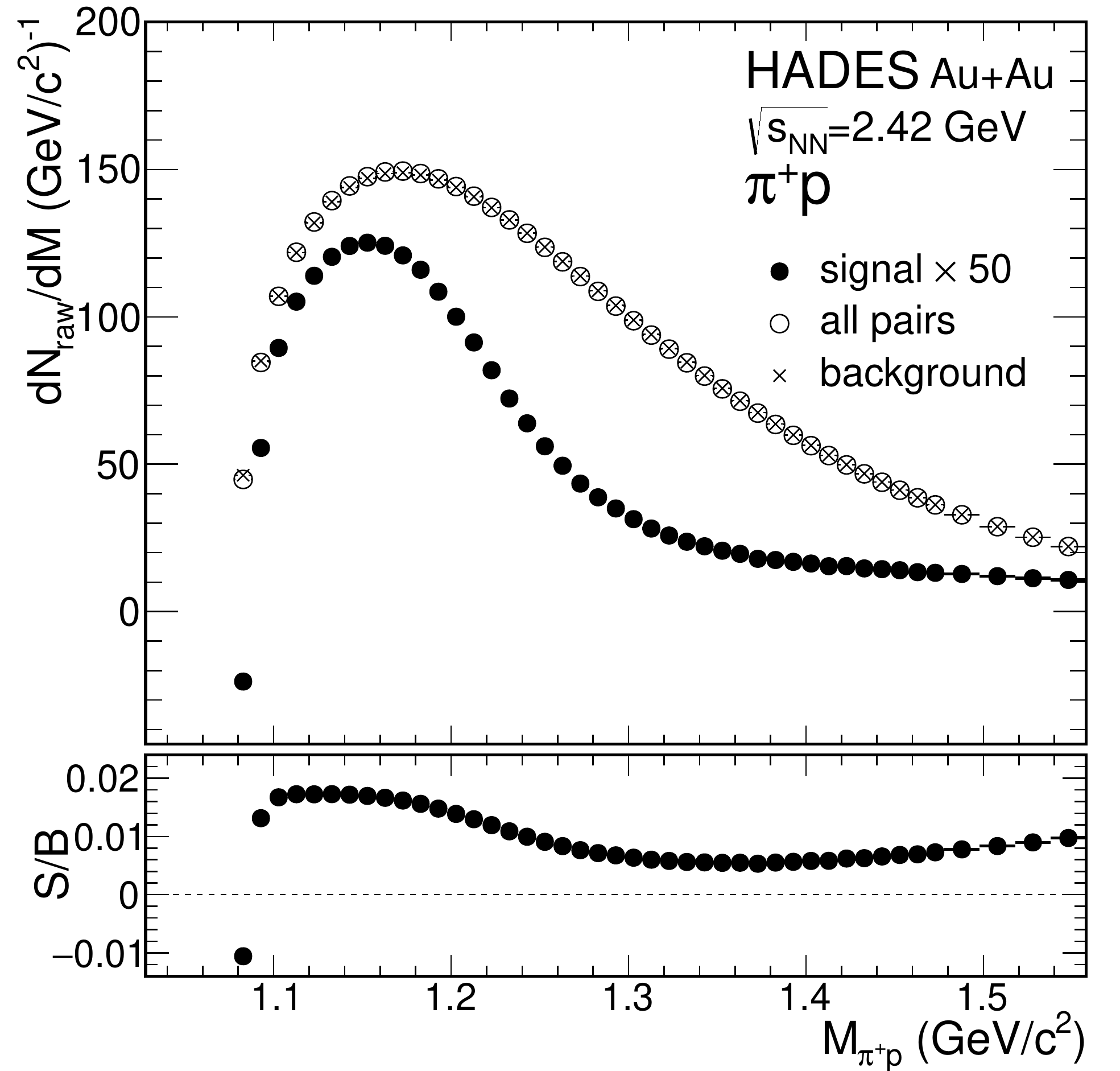}\caption{Top panel: Uncorrected invariant-mass $M_{\uppi^+{\text p}}$ distributions:  reconstructed $\uppi^+$p pairs  (open circles),  calculated combinatorial background (crosses) and  the result, i.e.~"signal", obtained by the subtraction of the background from the distribution of all measured pairs (full circles), scaled by a factor of 50. Bottom panel: Signal-to-background ratio as a function of $M_{\uppi^+{\text{p}}}$.}\label{fig_raw_sig}
\end{figure}

\begin{figure}[tb!]
\centering
\includegraphics[width=0.85\linewidth]{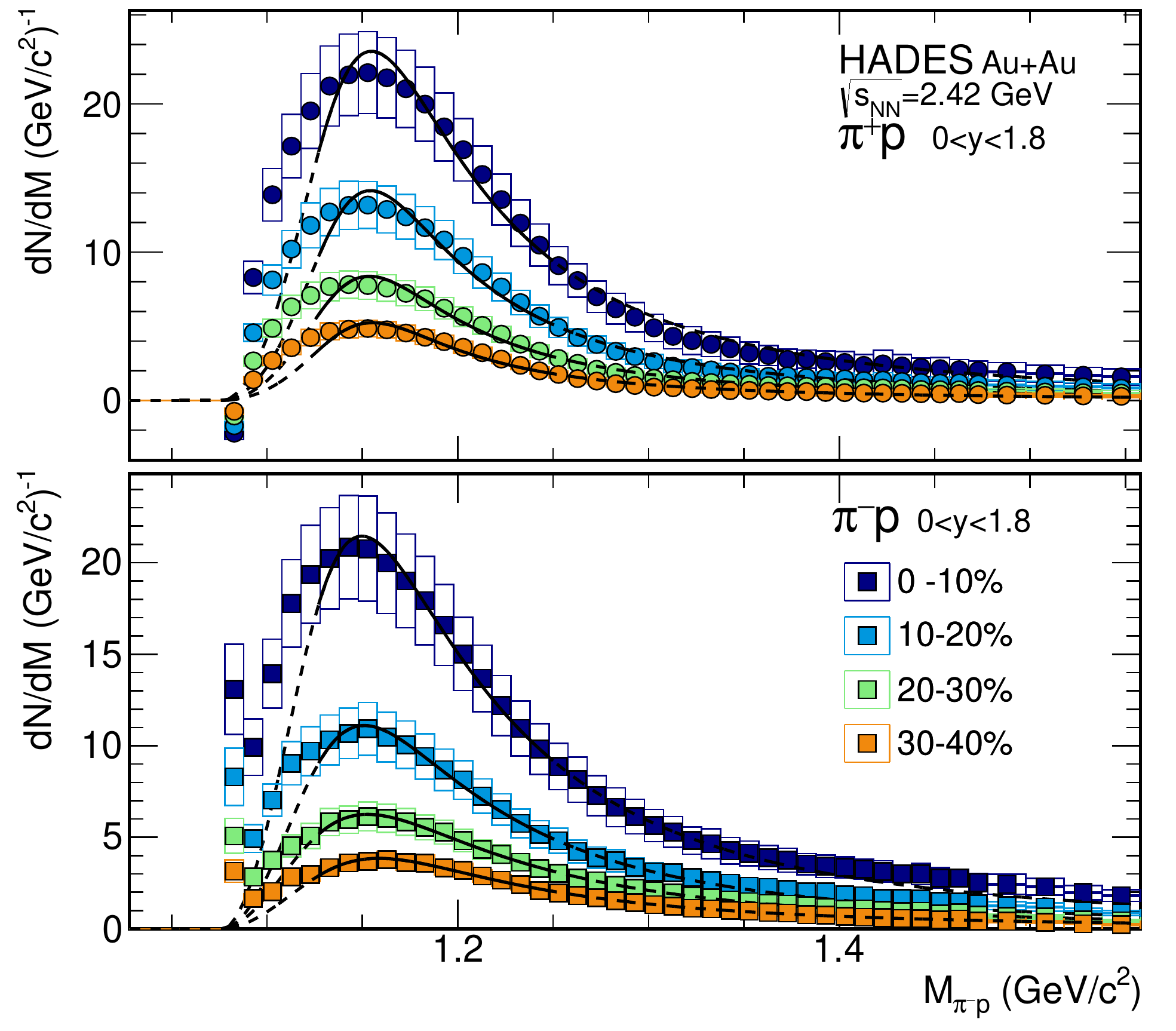}\label{fig_total_pip_p}
\caption{Efficiency and acceptance-corrected invariant mass distribution of correlated $\uppi^+$p (top panel) and $\uppi^-$p (bottom panel) pairs for four centrality intervals within the rapidity window of  $0<y<1.8$. 
The solid curve is a fit with formula~(\ref{formula_cugnon}) and the dashed curve is its continuation outside the fitting range. The boxes represent the systematic errors of the measurement whereas the statistical uncertainties are smaller than sizes of markers.}\label{fig_total_pi_p}
\end{figure}

\begin{figure}[tb!]
\centering
\includegraphics[width=0.85\linewidth]{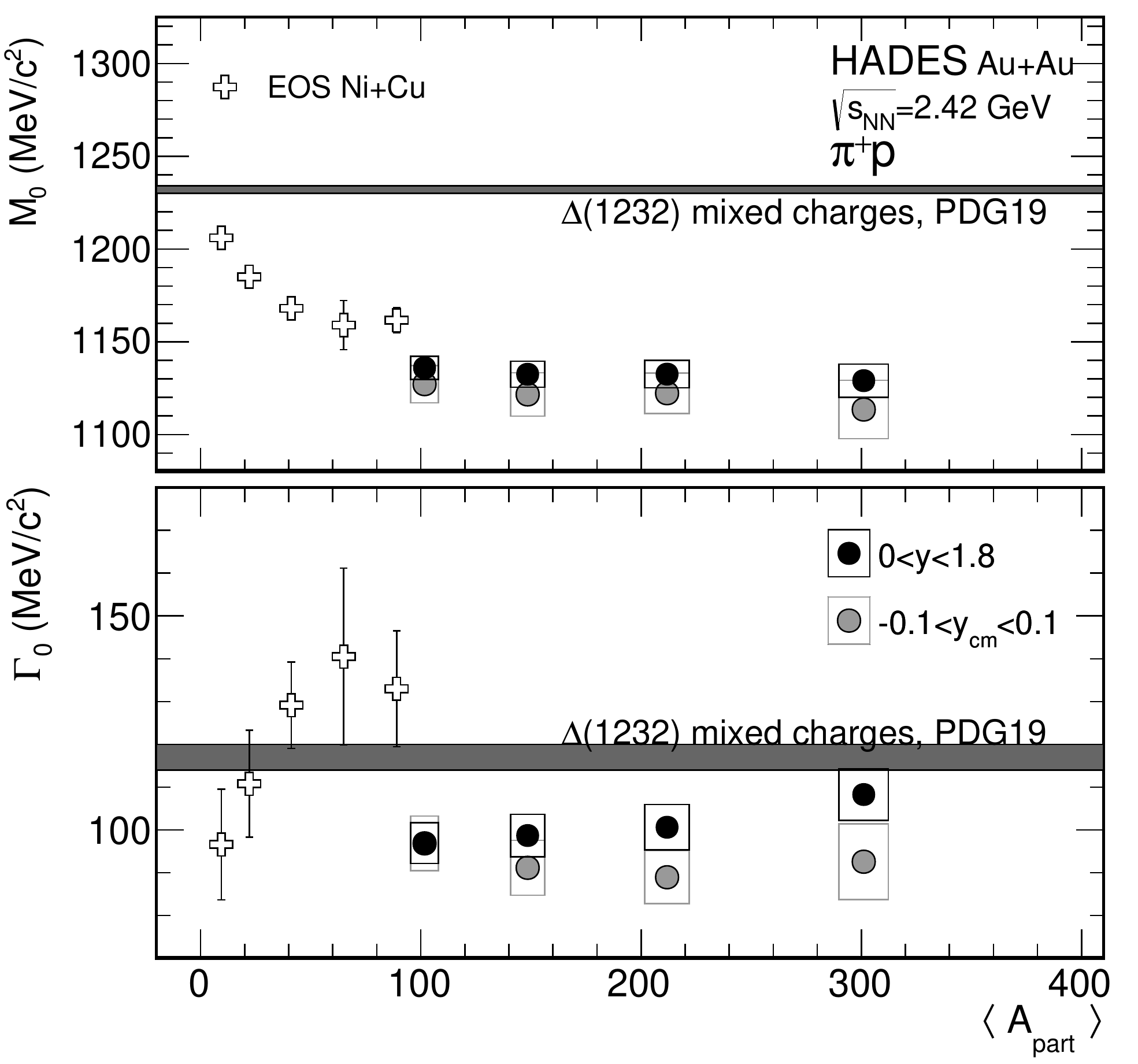}\label{fig_bw_mass}
\caption{Top panel: 
Mass parameter $M_0$ obtained by a fit with
Eq. (\ref{formula_cugnon}) for 
$\uppi^{+}$p pairs as a function of the mean number of participants measured in Au+Au (full cirlces) compared to the results obtained with the same procedure from Ni+Cu at $\sqrt{s_{NN}}$=2.69 GeV measured by EOS-TPC (open crosses)~\cite{EOS_Delta_NiCu_Hjort:1997}. 
The original EOS-TPC points are given as a function of the impact parameter; they have been recalculated to the mean number of participants using a geometrical model~\cite{Eskola:1988yh}. 
Bottom panel: Width parameter $\Gamma_0$ according to Eq.  (\ref{formula_cugnon}) of the same distributions. 
Only the systematic errors are shown as boxes, as the size of statistical uncertainties is negligible and smaller than the size of the markers.}
\label{fig_bw_mass_width}
\end{figure}

A gold beam of 1.23\textit{A} GeV delivered by the SIS18 synchrotron collided with a thin (1.4~\% interaction probability) segmented gold target placed in the HADES set-up which was used to detect $\uppi^+, \uppi^-$ and protons used in this analysis~\cite{HADES_setup:2009}. 
The magnetic field generated by six identical superconducting coils deflected charged particle trajectories. Momenta were reconstructed with a precision better than 2~\%, using four stations of low-mass drift chambers (MDC), two in front and two behind the magnet.
The time-of-flight (TOF) was measured by a diamond START detector~\cite{START} located in front of the gold target and by two STOP systems after the MDC stations. 
The angles $15<\theta<45^{\circ}$ are covered by a highly segmented Resistive Plate Chamber detector~\cite{Kornakov:2014cua}. 
The region corresponding to $45<\theta<85^{\circ}$ is covered by a plastic scintillator TOF Wall~\cite{Agodi:Hades_TOF_2002}. 
Pions and protons have been identified using a 2.5$\sigma$ selection criterion of a combined measurement of TOF and specific energy loss in the tracking detectors.
Their four-momenta are obtained assuming vacuum masses of identified particles.   
The geometrical acceptance of the detectors covers about 85~\% of full azimuth and the polar angle region between 15 and 85 degrees, which is equivalent to a rapidity coverage of $0<y<1.8$ in the laboratory frame (distributed symmetrically around mid-rapidity, $y_{\rm cm}=0.74$). 
An online multiplicity trigger based on the number of hits counted in the TOF Wall detector selected the 43~\% most central collisions.  
Further off-line event centrality selection was based on dedicated Glauber Monte-Carlo calculations  \cite{HADES:Centrality2017}. Four centrality bins corresponding to 10~\% changes in
the differential cross section (0--10~\%, 10--20~\%, 20--30~\%, 30--40~\%) containing about 5x10$^8$ events each were defined. 
Pions (protons) with laboratory momenta above 65~MeV/c (350~MeV/c) have been reconstructed. 
The particle purity after identification is better than 95~\%. 
Further details of the detector setup, tracking and particle reconstruction can be found in \cite{HADES_setup:2009}.

\begin{figure}[tbh]
\centering
\includegraphics[width=0.99\linewidth]{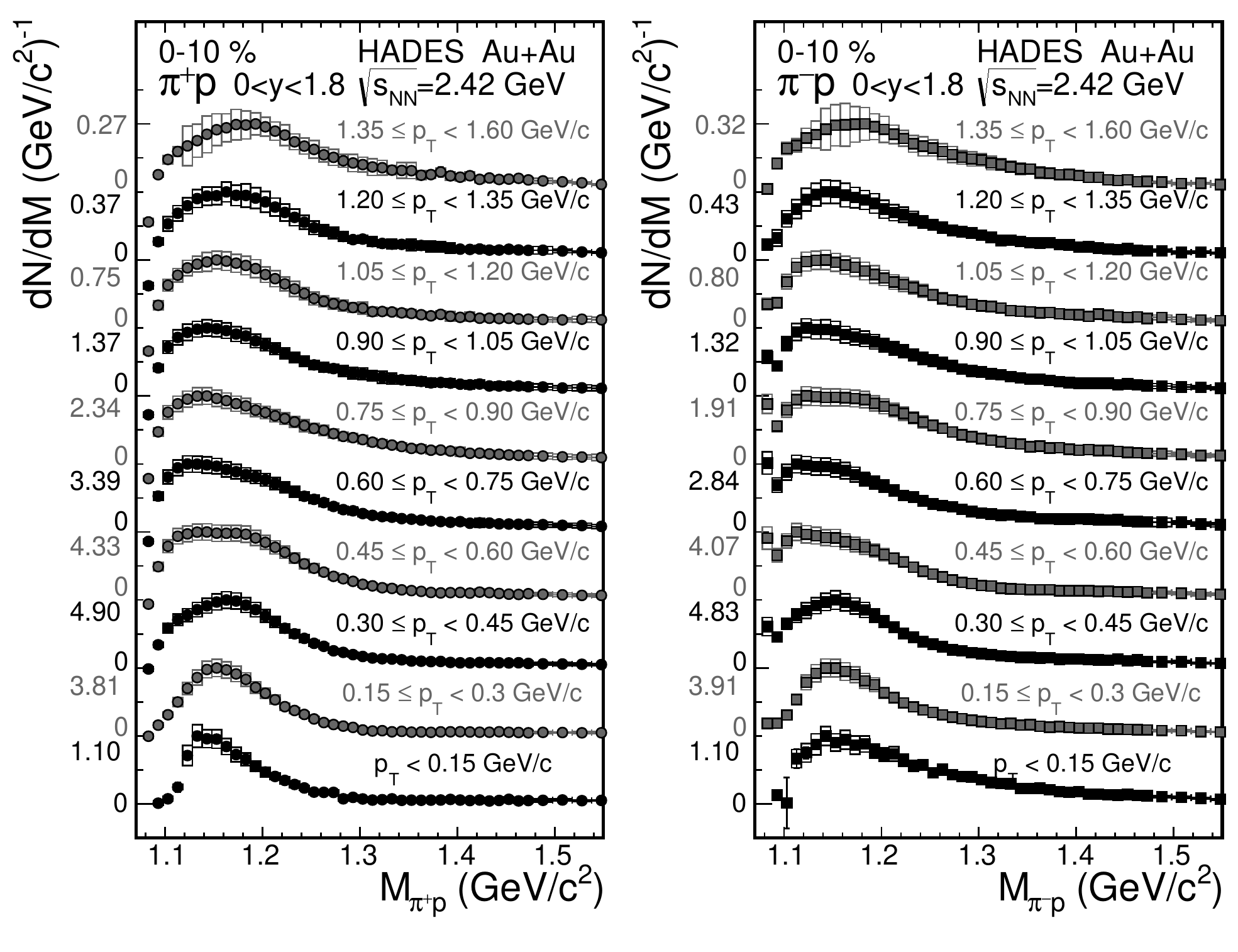}
\caption{Reconstructed invariant-mass distribution of $\uppi^+$p (left column) and $\uppi^-$p (right column) pairs for 10 subsequent $p_{\text{T}}$ intervals. The boxes depict the systematic errors of the measurement.}
\label{fig_mass_diff_pt}
\end{figure}

\begin{figure}[tbh]
\centering
\includegraphics[width=0.85\linewidth]{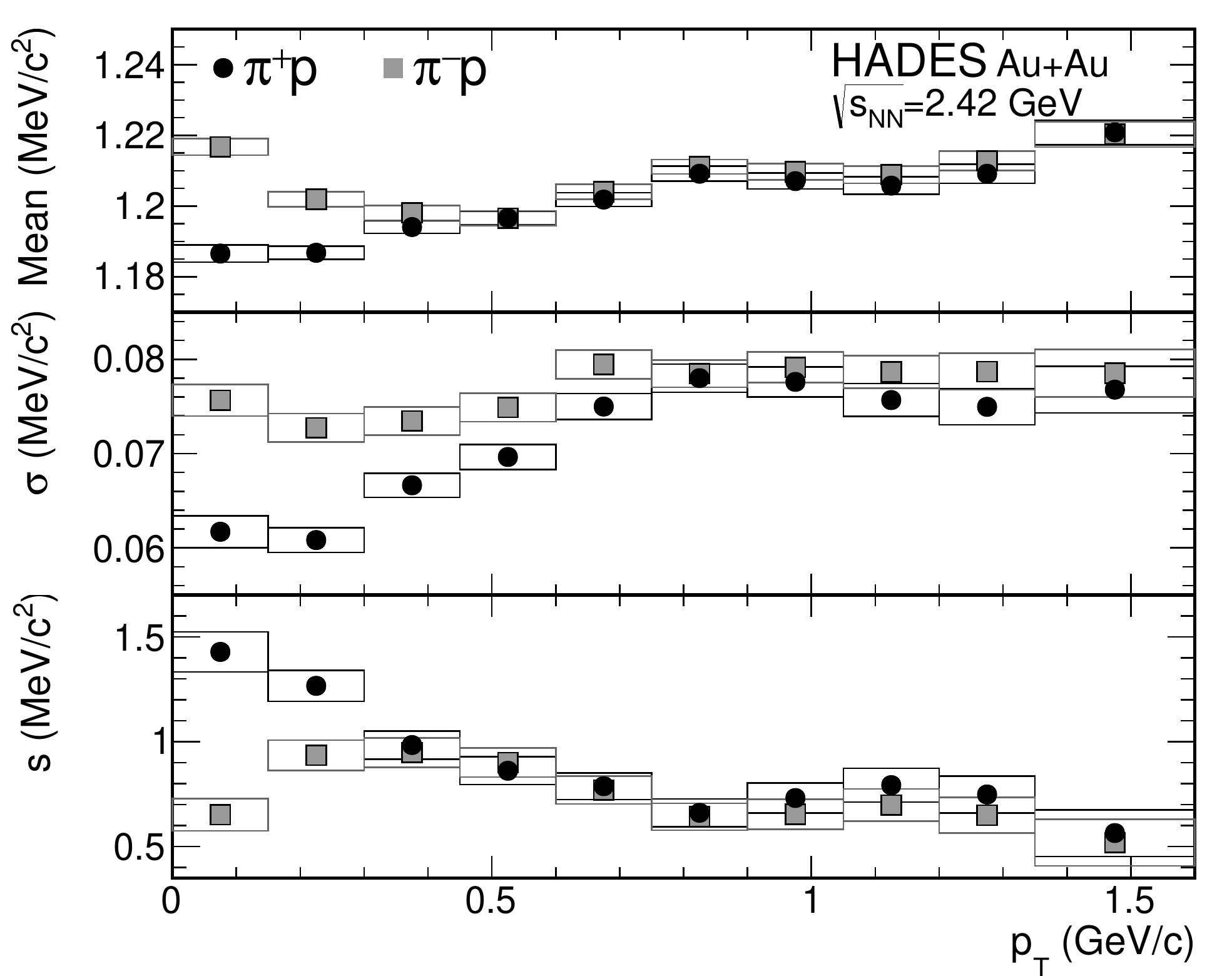}
\caption{Mean (top panel), standard deviation $\sigma$ (middle panel) and skewness (bottom panel) of the $p_{\rm T}$ distributions presented in Fig. \ref{fig_mass_diff_pt} shown as circles for $\uppi^+$p channel and as squares for $\uppi^-$p channel.
The boxes depict the systematic errors of the measurement.}
\label{fig_mass_mean_var_sk}
\end{figure}

\begin{figure}[tbh]
\centering
\includegraphics[width=0.85\linewidth]{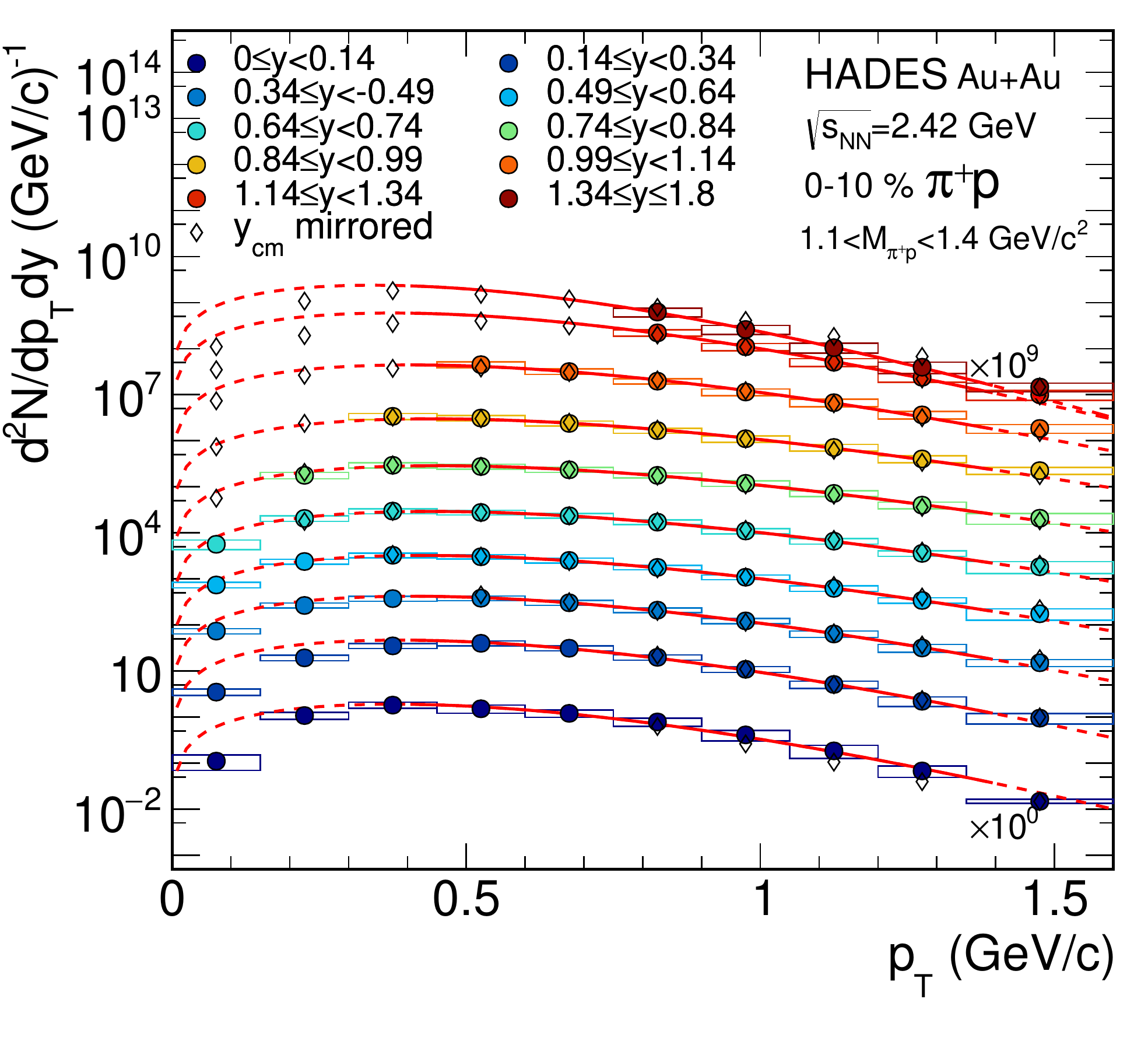}
\caption{Transverse momentum distribution for the 0-10~\% centrality class of $\uppi^+$p pairs with invariant mass between 1.1 and 1.4~GeV$/c^2$ for 10 rapidity intervals shown by coloured full circles and the $y_{cm}$-mirrored by diamonds.
Solid curves are fits with formula (\ref{eq:bessel}) to the measured data, and the dashed curves indicate the extrapolation outside the fit region.
The boxes depict the systematic errors of the measurement.}
\label{fig_yield_diff_pt}
\end{figure}

\begin{figure}[tbh]
\centering
\includegraphics[width=0.85\linewidth]{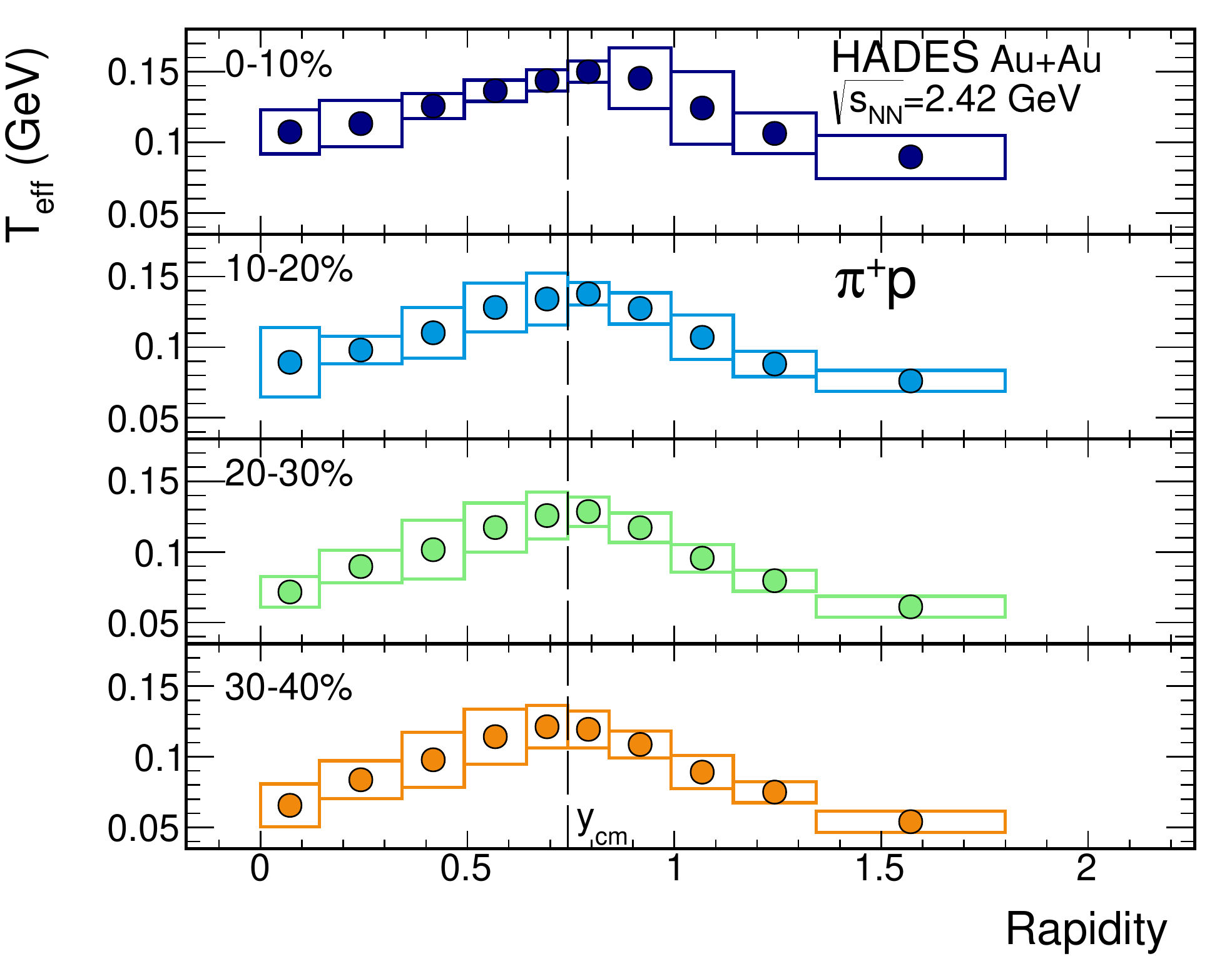}
\caption{Effective temperature T$_{\rm eff}$ obtained from fits of Eq.   (\ref{eq:bessel}) to the transverse momentum distribution of $\uppi^+$p pairs with invariant masses between 1.1 and 1.4~GeV$/c^2$.
The long-dashed line shows the center-of-mass rapidity $y_{\rm cm}= 0.74$.}
\label{fig_inverse_slope_t_cent}
\end{figure}

\begin{figure}[htb]
\centering
\includegraphics[width=0.9\linewidth]{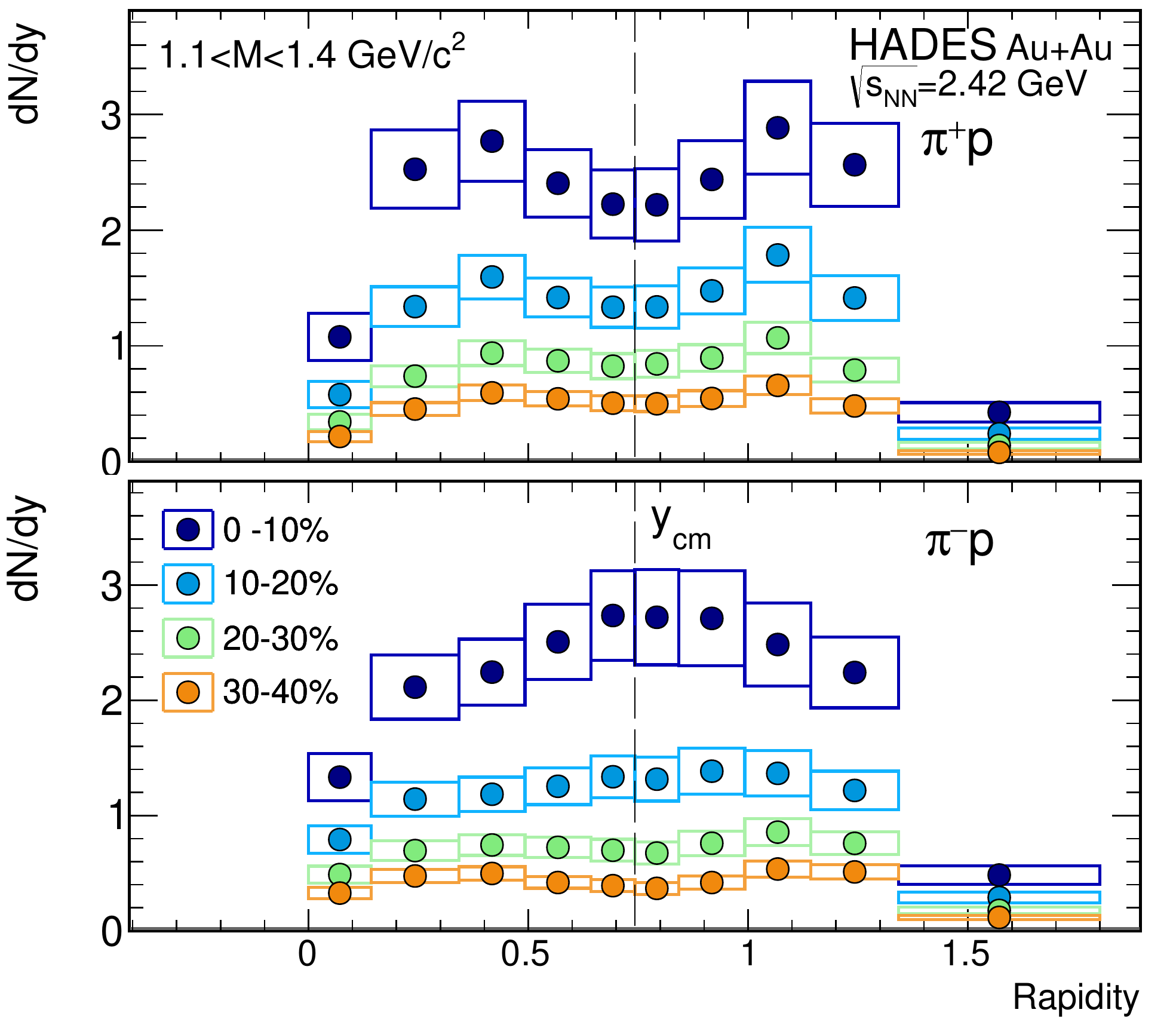}
\label{fig_rapidity_pip_p}\label{fig_rapidity_pim_p}
\caption{Rapidity density distributions of correlated $\uppi^+$p (top panel) and $\uppi^-$p (bottom panel) pairs integrated within the $1.1<M_{\text{inv}}~($GeV$/c^2) < 1.4$ range for four centrality classes.
The boxes represent the systematic errors of the measurement.
The long-dashed lines represent the mid-rapidity.}
\label{fig_rapidity}
\end{figure}

\begin{figure}[htb]
\centering
\includegraphics[width=0.9\linewidth]{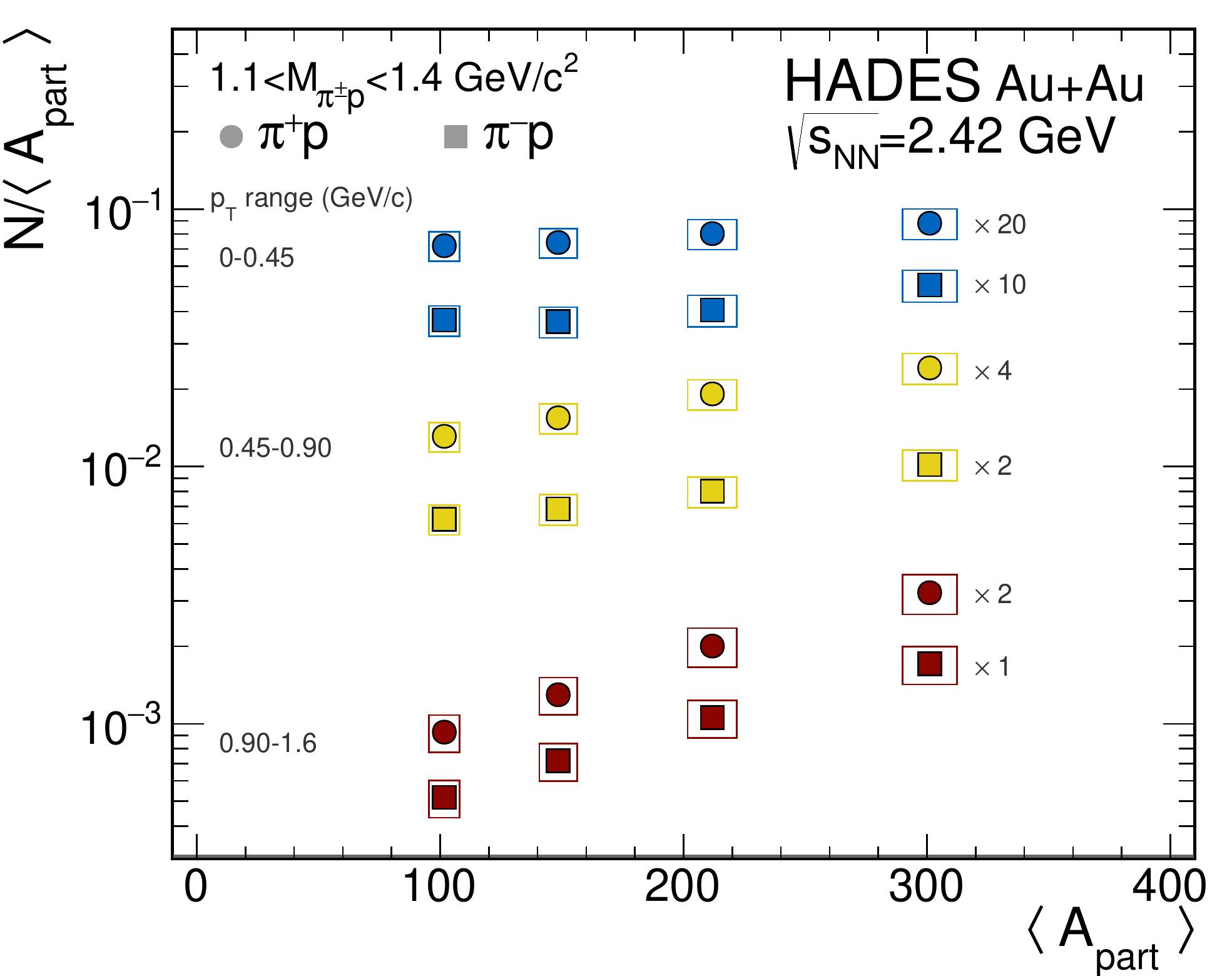}
\caption{Number of correlated pairs N per event and per mean number of participants $<$Apart$>$ of $\uppi^+$p (filled circles) and $\uppi^-$p (filled squares) pairs for three different $p_{\text{T}}$ ranges extrapolated to full rapidity as a function of $<$Apart$>$.
The boxes represent the systematic errors of the measurement.}
\label{fig_yield_mass_differential_apart}
\end{figure}

The correlated signal of $\uppi^{\pm}$p pairs is obtained by combining event by event all of the reconstructed and identified tracks available in the data sample. 
The distributions of pairs are binned as follows: 150 MeV$/c$ wide bins in $p_{\text{T}}$  from 0 to 1350~MeV$/c$ and one additional 250 MeV$/c$ wide bin up to 1600~MeV$/c$, 10 bins in laboratory rapidity from 0 to 1.8 with increasing bin size, 50 bins in pair invariant mass ($M_{\text{inv}}$) from the  $\uppi$p threshold to 1900 MeV$/c^2$, 10 equal bins from -1 to 1 in $\cos\theta$, where $\theta$ is the angle of the pion in the rest frame of the pair with respect to the motion of the pair in the center-of-mass frame of the collision. 
The background is obtained using a procedure which splits iteratively the total measured sample in correlated and uncorrelated pion-proton pairs  using the random track rotation technique described in details in~\cite{Kornakov:iterative_2018}. 
The procedure accounts for the detector pair acceptance and efficiency and a correlation of tracks with respect to the orientation of the event plane. 
The invariant-mass spectra of $\uppi^{+}$p pairs before and after the combinatorial background subtraction are shown in Fig.~\ref{fig_raw_sig} together with the signal-to-background ratio which ranges from 0.005 to 0.02.
Note that the first bin at the threshold invariant mass has a negative value which we attribute to the dominant repulsive Coulomb interaction between the two identically charged particles. 

To calculate efficiency corrections, detailed GEANT3 Monte-Carlo~\cite{Brun:1987ma} simulations have been performed using a generator based on the UrQMD transport  model~\cite{Bass:UrQMD_1998}.
The mean  efficiency for pion-proton pair reconstruction is of the order of 0.5. 
Acceptance corrections are based on a resonance cocktail produced with the Pluto event generator~\cite{Frohlich:2009eu} consisting of a thermal distribution with T=65 MeV and blast velocity of $\beta=0.3$ of $\Delta(1232)$, N$(1440)$, N$(1520)$/N$(1535)$ and $\Delta(1600)$ with a relative abundance of 5:1:0.5:0.5 and branching ratios according to \cite{PDG:2016}.
The model input values have been inspired by thermal models~\cite{Andronic:2005yp,Cleymans:2005xv}, transport calculations~\cite{Reichert:2020yhx} and measurements~\cite{Agakishiev:2010rs} for colliding systems of similar size and energy. 
The relative populations of the resonances have been adjusted by data driven multiplicities in the covered phase space.   
The pair acceptance varies between  0.2 and 0.8, depending mainly on rapidity with negligible impact on the shape of the reconstructed distribution.  
Different cocktail compositions consisting $e.g.$ of only $\Delta(1232)$ or $N(1440)$ as well as fireball temperatures varying from 35 to 50 MeV have been considered for the evaluation of systematic uncertainties. 
Taking advantage of the symmetry of the colliding system, pairs with low transverse momentum at forward rapidity, which lay outside the detector acceptance, have been estimated from the backward rapidities covered by the detector.
The extrapolation to full rapidity has been performed using a Gaussian function centered at mid-rapidity, the extrapolated yields are not more than 20~\% of the total. 
Main sources of systematic uncertainties attributed to the signal yield are due to the estimation of the background,  acceptance, efficiency corrections, and particle identification.
These four sources of systematic errors contribute with 10~\%, 10--20~\%, 5~\%, and  4--10~\%, respectively, and their contributions, considered as being independent, are quadratically added. 
The total systematic uncertainty is 15~\% below 1.3 GeV$/c^2$ and 25--30~\% for larger invariant masses.
Statistical errors are negligible with respect to the systematic uncertainties. 
The yields measured at forward and backward rapidities are found to be consistent within systematic uncertainties.

\section{Results}

Efficiency and acceptance corrected mass distributions of correlated $\uppi^{\pm}$p pairs are shown in Fig.~\ref{fig_total_pi_p} for four centrality classes within the HADES rapidity coverage. 
These distributions show a resonant $\Delta(1232)$ peak with a maximum around 1.15 GeV/c$^2$.  
The suppression and enhancement at threshold are in part due to the repulsive and attractive Coulomb effect, but also due to
$\uppi^-$N s-wave scattering as in the isospin 1/2 channel the scattering phase shift is attractive, whereas in the isospin 3/2 channel is repulsive \cite{Weinhold_thermo_delta:1997,Weinhold_thermo_phd:1998,Kisiel:2009eh}.
One should also note, that the $\Lambda$ signal  is not  visible in the $\uppi^-$p channel due to both the broad mass binning (width of the reconstructed $\Lambda$ is about 2.5 MeV$/c^2$) and its low production cross section below the free nucleon-nucleon threshold \cite{HADES_Lambda_k0_auau:2018}.

In order to compare to previous measurements, the resonant $\uppi $p cross section can be parametrized according to \cite{Cugnon_medium_effects_in_pion:1988} as follows:
\begin{equation}
    \sigma_{BW}(M_{\uppi{\text {p}}})=\frac{q^3}{q^3+\mu^3}\frac{\sigma_0}{1+4[(M_{\uppi{\text {p}}}-M_0)/\Gamma_0]^2},\label{formula_cugnon}
\end{equation}
where $q$ is the momentum in the pair center-of-mass frame, $\mu=180$~MeV/c, $M_{\uppi{\text {p}}}$ is the pair invariant mass, $M_0$ and $\Gamma_0$ are the Breit-Wigner resonance mass and width parameters, respectively, and $\sigma_0$ is a free normalization factor. 
The values extracted with this parametrization for a free $\Delta$ resonance with Breit-Wigner mass distribution according to PDG ($\text{M}_0=1232~\text{GeV}/c^2$ and $\Gamma_0=117~\text{GeV}/c^2$)~\cite{PDG:2016} have been found to be 1215~MeV$/c^2$ and 110~MeV$/c^2$ for the mass and width~\cite{DIOGENE_Trzaska:1991}, respectively. 
Fits performed with Eq~\ref{formula_cugnon} in the invariant-mass range from 1125~MeV$/c^2$ to 1250~MeV$/c^2$ are shown in Fig.~\ref{fig_total_pi_p} as solid black curves and the extrapolation of the function to the low- and high-mass regions as dashed curves. 
The deviations from the formula in these regions are expected due to the long-range Coulomb interaction between the decay products and the only partially reconstructed three-body decays ($e.g.$ $\uppi \textrm{N} \rightarrow \uppi\uppi \textrm{N}$) of higher-lying resonances and non-resonant contributions. 
The yield of masses greater than 1.25 GeV$/c^2$ is larger for the $\uppi^-$p as expected due to the contribution of the N$(1440)$ resonance which does not populate the $\uppi^+$p channel. 

The  mass ($M_0$) and width ($\Gamma_0$) parameters extracted employing Eq.~(\ref{formula_cugnon}) for $\uppi^+$p pairs are presented in  Fig.~\ref{fig_bw_mass_width} as a function of the $\langle\mathrm{A}_{\mathrm{part}}\rangle$ value of each centrality class~\cite{HADES:Centrality2017}. 
We give the $\uppi^+{\rm p}$ channel some preference with the rationale that the $P_{33}$(1232) resonance is much more prominent in the cross section $\uppi^+{\rm p}\rightarrow {\rm X}$ than in the cross section $\uppi^-{\rm p}\rightarrow {\rm X}$, where also more resonances contribute~\cite{EOS_Delta_NiCu_Hjort:1997}.
The $M_0$ values deduced for the invariant mass spectra integrated over the whole available rapidity range ($0 < y < 1.8$) are lower by more than 85~MeV$/c^2$ than the free values.   
The $\Gamma_0$ values are reduced by 20 to 25 MeV$/c^2$. 
No strong dependence on the centrality of the collision can be observed in HADES data (see Fig.~\ref{fig_bw_mass_width} black circles). 
Values obtained in a narrow bin around the center of mass rapidity ($y_{\text{cm}} \pm 0.1$) are systematically lower (see Fig.~\ref{fig_bw_mass_width} gray circles). 
Previous results on $\Delta$ mass and width reconstructed by the DIOGENE experiment in proton+C/Nb/Pb reactions~\cite{DIOGENE_Trzaska:1991} reported a significant reduction (10 -- 60 MeV$/c^2$) of  $M_0$ with respect to the free value for all three systems. 
The EOS-TPC experiment provided measurements for the Ni+Cu system at 1.97$A$ GeV~\cite{EOS_Delta_NiCu_Hjort:1997} showing a decrease of the $\Delta$ mass values with $\langle\mathrm{A}_{\mathrm{part}}\rangle$, for $\langle\mathrm{A}_{\mathrm{part}}\rangle$ values up to 90, as also shown in fig.~\ref{fig_bw_mass_width}. 
The $M_0$ values from our experiment are consistent with the previous measurements and show  a possible saturation of the mass decrease for $\langle\mathrm{A}_{\mathrm{part}}\rangle$ larger than 100. 
The widths extracted in former experiments did not provide a conclusive picture, while our measurements clearly exhibit  values of $\Gamma_0$ lower than the free value.

The invariant-mass differential spectra for several $p_{\text{T}}$ bins of $\uppi^+$p and $\uppi^-$p pairs are shown in Fig.~\ref{fig_mass_diff_pt}. 
In order to quantify the dependence of the shape of the invariant mass distributions  on the transverse pair momentum, we have extracted the first, second and third moments of the spectra in an invariant-mass range between 1.1 and 1.4 GeV/c$^2$.  
Fig.~\ref{fig_mass_mean_var_sk} depicts the extracted mean, sigma and skewness as a function of the pair transverse momentum $p_{\text{T}}$, which show  a significant dependence on $p_{\text{T}}$. 
For transverse momenta above 0.5 GeV/c, the extracted parameters have the same values within errors for  pion-proton correlated pairs of both charges. 
The mean and standard deviation are slowly rising, whereas the skewness drops, as $p_{\text{T}}$ increases. It is noticeable that, even for the highest $p_{\text{T}}$, the mass parameter is still far from the vacuum value.
For transverse momenta below 0.5 GeV/c a significant difference of the spectral shape between $\uppi^-$p and $\uppi^+$p becomes visible. 
For comparison, the corresponding moments of the invariant-mass distribution of the $\Delta$ produced in a pp collision~\cite{Agakishiev:2017nxc} at the same center-of-mass energy are 1.24 GeV$/c^2$, 0.057 GeV$/c^2$ and 0.5, respectively.
The rise of the most probable mass as a function of pair $p_{\text{T}}$ in Au+Au is in accordance with model predictions \cite{Bass:high_pt_pions_probes_1994,Bass:probin_delta_matter_AuAu_1gev_1994,Vogel_high_pt_probes:2009,Reichert2019_UrQMD_Delta_For_HADES}, where the effect is related to the time evolution of the mass of the resonance and the decoupling from the medium. 
Indeed, the mean mass of the $\Delta$ resonance is expected to decrease as a function of time, due to the decreasing energy available in subsequent NN and $\uppi$N collisions. 
The majority of low-$p_{\text{T}}$ resonances decouple from the medium only at later stages, close to freeze-out, when the available energy in the $\uppi$N collision is reduced.  
In this scenario, high-$p_{\text{T}}$ $\Delta$ resonances, which are excited and may decouple earlier from the medium, are less affected by these phase space effects.

The  $p_{\text{T}}$-differential yield of $\uppi^{+}$p pairs with invariant mass between 1.1 and 1.4 GeV$/c^2$ for 10 rapidity ranges are shown in Fig.~\ref{fig_yield_diff_pt}. 
The measurement has full coverage in the backward rapidity region, but some $p_{\text{T}}$ bins are missing for forward rapidities. 
The rapidity-mirrored yields are shown as empty squares.
They are found to be consistent within uncertainties to the expected values.
These $p_{\text{T}}$ distributions have been fitted using the function
\begin{equation}
    \frac{1}{p_{\text{T}}}\frac{\text{d}N}{\text{d}p_\text{T}} \propto m_{\text{T}}\text{K}_1\left(\frac{m_\text{T}}{T_{\rm eff}}\right),\label{eq:bessel}
\end{equation}
where $m_{\text{T}}=\sqrt{M^2+p_{\text{T}}^2}$  is the pair transverse mass, $M=1.2$~GeV$/c^2$ and $\text{K}_1$ is the modified Bessel function of the second kind, and $T_{\rm eff}$ is the inverse slope parameter which characterizes the source temperature and additional effects originating from the collective motion and decays. 
Therefore, we refer to it as the effective temperature.

The effective temperatures obtained from the fits with Eq.~(\ref{eq:bessel}) of the $p_{\text{T}}$ spectra of $\uppi^{+}$p pairs are shown in  Fig.~\ref{fig_inverse_slope_t_cent} as a function of the rapidity for the four centrality classes.
The maximum $T_{\rm eff}$ for the 0--10 \% event centrality selection yields 150 MeV and is reached around $y_{\text{cm}}$. 
The effective temperature decreases in the peripheral collisions by 10 MeV for each 10~\% of event centrality, 
reaching 120 MeV for 30--40~\% event centrality at mid-rapidity. 
Considering a radial-blast expansion of the system with a common expansion velocity of $\beta=0.3$ \cite{Galatyuk:2015pkq,Szala:ect19}, the $\Delta^{++}$ freeze-out temperature is $T_{\rm fo}\simeq50$~MeV in central collisions.
This value points to a late decoupling of the resonances in comparison to the value of $71.8 \pm 2.1$~MeV obtained from the di-lepton invariant mass spectra~\cite{HadesNature2019}.     

In Fig.~\ref{fig_rapidity} the $p_{\text{T}}$ and mass $1.1<M_{\text{inv}} $~GeV$/c^2 < 1.4$ integrated rapidity distributions are shown for both channels. 
Significant differences can be observed for the most central collisions.
Similar distributions as a function of rapidity have been reported for the mean mass in theoretical studies with UrQMD of Au+Au collisions at 1.23$A$~GeV  and interpreted as a long-evolving ${\rm{\Delta }}\leftrightarrow \uppi +\text{N}$ cycle, the effect being most prominent at mid-rapidity for pairs with $p_{\text{T}}$ smaller than 0.75 GeV/c~\cite{Reichert2019_UrQMD_Delta_For_HADES}. However, the   observed isospin-dependence in our data has not been studied yet. 

\begin{table}[htb!]
    \centering
     \caption{Exponents from the 
     $\langle\mathrm{A}_{\mathrm{part}}\rangle$ scaling, obtained from the yields as a function of centrality for $\uppi^{\pm}$p pairs in the $\Delta$ mass region (1.1--1.4 GeV$/c^2$) for the $p_{\text{T}}$-integrated data in three intervals as shown in  Fig.~\ref{fig_yield_mass_differential_apart}.}
    \small\addtolength{\tabcolsep}{-2pt}
    \begin{tabular}{l|c|c}
         $p_{\text{T}}$ (GeV$/c$) & $\alpha_{\Delta^{++}}$ &$\alpha_{\Delta^0}$ \\
         \hline
         0 -- 1.6 &$1.49\pm0.08^{\text{st}}\pm0.21^{\text{sy}}$ &$1.48\pm0.09^{\text{st}}\pm0.20^{\text{sy}}$\\
         \hline
         0 -- 0.45 &$1.20\pm0.09^{\text{st}}\pm0.19^{\text{sy}}$
         &$1.33\pm0.09^{\text{st}}\pm0.21^{\text{sy}}$\\
         0.45 -- 0.9
         &$1.58\pm0.11^{\text{st}}\pm0.20^{\text{sy}}$
         &$1.47\pm0.10^{\text{st}}\pm0.20^{\text{sy}}$\\
         0.9 -- 1.6 &$2.18\pm0.14^{\text{st}}\pm0.26^{\text{sy}}$ &$2.13\pm0.14^{\text{st}}\pm0.25^{\text{sy}}$
    \end{tabular}
    \label{tab:alpha_pars}
\end{table}

\section{Discussion of centrality dependence}

The number of correlated pairs with invariant masses $1.1<M_{\text{inv}}~$GeV$/c^2 < 1.4$ per event normalized by the number of participating nucleons is shown as a function of the number of participating nucleons per event in Fig.~\ref{fig_yield_mass_differential_apart} for three different intervals of pair transverse momentum. 
While the pairs with $p_{\text{T}}<0.45$ GeV$/c$ have an almost flat distribution as a function of the mean number of participants, pairs with $0.45\leq p_{\text{T}}<0.9$ GeV$/c$ have already a significant excess of yield for central events. 
This trend is continued for the pairs with $p_{\text{T}}>0.9$ GeV$/c$, where an even stronger increase of the yield is observed towards central collisions. 
Quantitatively, both the neutral and double-charged pairs multiplicities can be described by a power-law function with free normalization
\begin{equation}
N \propto \langle\mathrm{A}_{\mathrm{part}}\rangle^{\alpha},\label{eq:scaling}
\end{equation}
where the $\alpha$ parameter value is $1.48\pm0.09$(stat)$\pm0.20$(syst) for $\uppi^-$p and $1.49\pm0.08$(stat)$\pm0.21$(syst) for $\uppi^+$p pairs,  respectively. 
The values obtained for the $\alpha$ parameter for both channels in the three transverse momentum intervals are summarized in Tab.~\ref{tab:alpha_pars}. 
Within errors both channels exhibit the same  $\langle\mathrm{A}_{\mathrm{part}}\rangle$ dependence. 
One of the explanations for the yield increase at large $p_{\text{T}}$ could be due to collective effects affecting those correlated pion-proton pairs which are expected to experience in-medium modifications~\cite{Larionov:2003av,Lenske:2018bgr}. 
This trend is at variance with the linear dependence of the inclusive pion production as a function of $\langle\mathrm{A}_{\mathrm{part}}\rangle$ observed in the same system~\cite{Adamczewski-Musch:2020vrg}.
The overall ratio between the $\uppi^-$p and $\uppi^{+}$p pairs is $0.98\pm0.2$ and is constant as a function of centrality within uncertainties. 
This number is larger than the value of 0.59 expected from excitations of $\Delta^{++}$ and $\Delta^0$ in first chance NN collisions in the different isospin channels and accounting for the Z/A ratio.  
This ratio is expected to be modified by absorption of the resonances or  successive interactions of pions with nucleons generating further $\Delta$ resonances, which both depend on isospin.
Considering also $\Delta^+$ and $\Delta^-$ excitation and decay, it can be deduced that twice as many $\uppi^-$ are produced as $\uppi^+$ in first chance NN collisions. 
The ratio of $\Delta^{++}$ to $\Delta^{0}$ formed in the subsequent steps is therefore expected to decrease. 
Scattering affects $\Delta^{++}$ much more than $\Delta^0$. 
Finally, charge-exchange processes lead to a reduction of $\Delta^{++}$, while $\Delta^0$ are produced and disappear with similar probabilities in such processes.  
Based on all these considerations, the ratio of  $\uppi^-$p to  $\uppi^+$p correlated pairs is expected to increase in the course of the collision, which is consistent with the observation of a similar number of correlated $\uppi^-$ and $\uppi^+$ pairs being detected. 
Although it includes all scattering and absorption effects, the calculations of~\cite{Reichert2019_UrQMD_Delta_For_HADES} predict a value of the ratio around 0.7, $i.e.$ intermediate between the single isobar model and our measurements. 
On the other hand, it has been shown for lower energy Au+Au collisions that a significant increase of the $\uppi^- / \uppi^+$ ratio is predicted by models including modified in-medium elastic and inelastic NN cross sections \cite{Yong:2010zg,Zhang:2017mps}.

\section{Summary}

The measured inclusive multi-differential spectra of correlated pion-proton pairs in Au+Au collisions at $\sqrt{s_{NN}}=2.42$~GeV have been reconstructed from the high statistics data sample recorded by HADES. 
We have performed the analysis using an iterative technique for the evaluation of the combinatorial background.
The data show a strong signal of the $\Delta(1232)$ resonance, as observed previously in other collision systems at similar beam energies. 

The centrality dependence of the mass and width parameters has been extracted from a fit of the pion-proton invariant mass distribution. 
In particular, the mass parameter exhibits a drop of 85 MeV$/c^2$ with respect to the $\Delta(1232)$ vacuum values, in line with EOS-TPC results~\cite{EOS_Delta_NiCu_Hjort:1997} for a lighter system Ni+Cu at $\sqrt{s_{NN}}=2.69$~GeV and DIOGENE results~\cite{DIOGENE_Trzaska:1991} for p+C/Nb/Pb at 0.8 and 1.6 GeV proton energy and the FOPI results~\cite{FOPI_Eskef:2001} as well.  
The analysis of the $p_{\text{T}}$-differential distributions result in a rapidity symmetric effective temperature profile with a maximum T = 150 MeV at mid-rapidity $y_{\rm cm}$=0.74 and lower by 50 MeV at the edges for the 0--10~\% event centrality class.
The local minimum of the yield at mid-rapidity and the variation of the maxima of the invariant mass as a function of $p_{\text{T}}$ point towards several ${\rm{\Delta }}\leftrightarrow \uppi +N$ cycles during the fireball lifetime. 
The scaling of the yield follows a power law function $N \propto \langle\mathrm{A}_{\mathrm{part}}\rangle^{\alpha}$ with $\alpha_{\Delta^{0}} = 1.48\pm0.09^{\text{st}}\pm0.20^{\text{sy}}$ for the neutral and $\alpha_{\Delta^{++}} =1.49\pm0.08^{\text{st}}\pm0.21^{\text{sy}}$ for doubly charged channels, respectively.  
The extracted spectra of correlated $\uppi$p pairs, their differential $p_{\text{T}}$ distributions and the observed structures in the rapidity  distributions provide a rich data base for a detailed comparison to different model calculations. 
Progress can be made regarding the temperature evolution at the late stages of the fireball evolution. 
The study of pion-nucleon correlated pairs
gives access to the collision dynamics, from the first chance NN collisions till the decoupling phase, which is an essential, yet not well known ingredient in hitherto model calculations.

We gratefully acknowledge support by the following grants:
SIP JUC Cracow, Cracow (Poland):
National Science Center, 2016/23/P/ST2/040 POLONEZ,
2017/25/N/ST2/00580,
2017/26/M/ST2/00600;
Warsaw University of Technology, Warsaw (Poland): OPUS grant from National Science Center of Poland 2017/27/B/ST2/01947;
TU Darmstadt, Darmstadt (Germany): VH-NG-823, DFG GRK 2128, DFG CRC-TR 211, BMBF:05P18RDFC1; 
Goethe-University, Frankfurt (Germany): BMBF:06FY9100I, BMBF:05P19RFFCA, GSI F$\&$E, HIC for FAIR (LOEWE); 
Goethe-University, Frankfurt (Germany) and TU Darmstadt, Darmstadt (Germany), ExtreMe Matter Institute EMMI at GSI Darmstadt; 
TU München, Garching (Germany): MLL München, DFG EClust 153, GSI TMLRG1316F, BmBF 05P15WOFCA, SFB 1258, DFG FAB898/2-2; 
the Russian Foundation for Basic Research (RFBR) funding within the research project no. 18-02-40086,
the National Research Nuclear University MEPhI in the framework of the Russian Academic Excellence Project (contract No.\textbackslash 02.a03.21.0005, 27.08.2013), the Ministry of Science and Higher Education of the Russian Federation, Project ``Fundamental properties of elementary particles and cosmology'' No 0723-2020-0041,
JLU Giessen, Giessen (Germany): BMBF:05P12RGGHM; 
IPN Orsay, Orsay Cedex (France): CNRS/IN2P3; 
NPI CAS, Rez, Rez (Czech Republic): MSMT LM2018112, OP VVV CZ.02.1.01/0.0/0.0/16 013/0001677, LTT17003.

\section*{References}
\bibliography{mybibfile}

\end{document}